\documentclass[letterpaper,11pt]{article}
\usepackage{jheppub}

\usepackage{braket}
\usepackage{tikz}
\usepackage{hyperref}
\usepackage{amssymb}
\usepackage{amsmath}
\usepackage{color}
\usepackage{esint}
\usepackage{soul}

\usepackage[utf8]{inputenc}

\DeclareMathOperator{\tr}{tr}
\DeclareMathOperator{\SU}{SU}
\DeclareMathOperator{\U}{U}

\begin{document}

\title{
Entanglement entropy and the large $N$ expansion \\
of two-dimensional Yang-Mills theory}

\abstract{
Two-dimensional Yang-Mills theory is a useful model of an exactly solvable gauge theory with a string theory dual at large $N$. We calculate entanglement entropy in the $1/N$ expansion by mapping the theory to a system of $N$ fermions interacting via a repulsive entropic force. The entropy is a sum of two terms: the ``Boltzmann entropy'', $\log \dim (R)$ per point of the entangling surface, which counts the number of distinct microstates, and the ``Shannon entropy'', $- \sum p_R \log p_R$, which captures fluctuations of the macroscopic state. We find that the entropy scales as $N^2$ in the large $N$ limit, and that at this order only the Boltzmann entropy contributes. We further show that the Shannon entropy scales linearly with $N$, and confirm this behaviour with numerical simulations. While the term of order $N$ is surprising from the point of view of the string dual --- in which only even powers of $N$ appear in the partition function --- we trace it to a breakdown of large $N$ counting caused by the replica trick. This mechanism could lead to corrections to holographic entanglement entropy larger than expected from semiclassical field theory.
}
\author[a]{William Donnelly,}
\author[a,b]{Sydney Timmerman,}
 \author[a,c]{and Nicolás Vald\'es-Meller}
\emailAdd{wdonnelly@perimeterinstitute.ca}
\emailAdd{stimmer2@jhu.edu}
\emailAdd{n.valdes.meller@gmail.com}
\affiliation[a]{Perimeter Institute for Theoretical Physics, 31 Caroline St. N, N2L 2Y5, Waterloo ON, Canada}
\affiliation[b]{Department of Physics and Astronomy, Johns Hopkins University, Charles Street, Baltimore, MD 21218, U.S.A.}
\affiliation[c]{Departamento de F\'isica, FCFM, Universidad de Chile, Blanco Encalada 2008, Santiago, Chile}

\maketitle

\newpage

\section{Introduction}

A longstanding challenge in quantum gravity is to understand the Bekenstein-Hawking entropy
\begin{equation} \label{Sbh}
    S_{\text{BH}} = \frac{A}{4 G}.
\end{equation}
It has been suggested that some or all of this entropy could come from entanglement \cite{Sorkin:2014kta,Bombelli:1986rw,Srednicki:1993im}: the entanglement entropy associated to a region of space also follows an area law, and diverges in the ultraviolet.
The entanglement entropy can be viewed as part of the one-loop correction to the Bekenstein-Hawking entropy, so it is most natural to consider the generalized entropy
\begin{equation} \label{Sgen}
    S_\text{gen} = \frac{\langle A \rangle}{4 G} + S_\text{out},
\end{equation}
where $S_\text{out}$ is the entropy of fields outside the horizon \cite{Susskind:1994sm,Jacobson:1994iw,Cooperman:2013iqr}.

In AdS/CFT, the relation between the Bekenstein-Hawking entropy and entanglement is made precise via the Ryu-Takayanagi formula and its subsequent generalizations \cite{Ryu:2006bv,Hubeny:2007xt,Faulkner:2013ana,Engelhardt:2014gca}.
The entanglement entropy of a holographic large $N$ conformal field theory has an expansion of the form
\begin{equation} \label{RT}
    S = \frac{\langle A \rangle}{4G} + S_\text{bulk} + O(G).
\end{equation}
Here the large $N$ expansion in the bulk becomes a small $G$ expansion, where $G \sim 1/N^2$ is Newton's constant in the bulk gravitational theory.
The leading term of order $N^2$ is the area of a classical extremal surface.
The first subleading correction appears at order $N^0$ and is given by the entanglement entropy of  bulk quantum fields across the minimal surface \cite{Faulkner:2013ana}.
Thus to understand the Bekenstein-Hawking entropy and its corrections, we should study entanglement entropy of gauge theories in the large $N$ expansion.

While entanglement of effective field theory on a fixed classical background captures the corrections to the Bekenstein-Hawking entropy, it does not provide a statistical explanation for the leading term in \eqref{Sgen}: the latter is simply a term in the effective action.
Unlike field theory, string theory could account for this term in the entropy \cite{Susskind:1993ws}.
In string theory the sphere diagram contributes to the entropy at order $g_s^{-2} \sim 1/G$, and its contribution can be understood as counting open string states with endpoints anchored to the entangling surface.
To calculate entanglement entropy directly within string theory remains fraught with challenging conceptual and technical issues \cite{Dabholkar:1994ai,He:2014gva,Balasubramanian:2018axm,Witten:2018xfj}, though in some cases it is feasible \cite{Hartnoll:2015fca,Hubeny:2019bje}. 

In this paper we will consider a particularly tractable example of large $N$ gauge/string duality: two-dimensional Yang-Mills theory.
Although Yang-Mills theory has no local degrees of freedom in two dimensions, it is a surprisingly rich theory.
On the sphere, the large $N$ theory has two phases:
at weak coupling it has a deconfined phase similar to random matrix theory and at strong coupling it has a description as a two-dimensional string theory \cite{Gross:1994mr}.
The two phases are separated by a third-order phase transition \cite{Douglas:1993iia}.
This is a useful model because it's exactly solvable,  we know how to calculate the entanglement entropy for any entangling surface \cite{Donnelly:2014gva}, and results are known to leading order in the $1/N$ expansion \cite{Gromov:2014kia}.
The entanglement entropy in this model has a description in the string theory, where it counts configurations of open strings ending on an \emph{entanglement brane} \cite{Donnelly:2016jet,Donnelly:2018ppr}.

Entanglement in gauge theory has some subtleties because the Hilbert space of physical states does not decompose as a usual tensor product \cite{Donnelly:2008vx,Buividovich:2008gq} (see \cite{Lin:2018bud} for a recent review).
Instead, the Hilbert space of a spatial region with boundary contains \emph{edge modes}.
Hilbert spaces are combined using an entangling product \cite{Donnelly:2016auv} which glues two regions along a shared boundary.

In two-dimensional Yang-Mills theory, this entangling product can be described quite explicitly \cite{Donnelly:2014gva,Donnelly:2016auv,Donnelly:2016jet}.
The Hilbert space of an interval is spanned by states $\ket{R,a,b}$ where $R$ is an irreducible representation of the gauge group $G$, and $a$ and $b$ are indices in the irreducible representation $R$ which live at the endpoints of the interval.
We can extend this to a general subsystem consisting of an arbitrary number of circles and intervals, whose boundary consists of $m$ points.
Any density matrix $\rho$ arising from a gauge-invariant state commutes with the action of the gauge group $G$ and takes the form $\rho = \bigoplus_R p_R \rho_R$, where $p_R$ is a probability distribution over the irreducible representations which specifies the state and $\rho_R$ is a maximally mixed state on all the degrees of freedom $a,b$.
The entropy is then a sum of two terms:\footnote{This definition extends naturally to lattice gauge theories in any dimension \cite{Donnelly:2011hn}. When the theory has local degrees of freedom there is a third term in \eqref{2terms} which captures their entropy --- for pure gauge theory in two dimensions, this term is absent.}
\begin{equation} \label{2terms}
S = S_\text{Boltzmann} + S_\text{Shannon}
\end{equation}
where
\begin{equation}
S_\text{Boltzmann} = m \sum_R p_R \log \dim(R), \qquad 
    S_\text{Shannon} = - \sum_{R} p_R \log(p_R).
\end{equation}
The Boltzmann entropy term is so called because it counts the number of indistinguishable states associated to the single ``macrostate'' labelled by $R$; a similar definition of quantum Boltzmann entropy was introduced in the context of black hole physics \cite{Wald:1979zz}.
The Shannon entropy, on the other hand, measures fluctuations of the gauge-invariant information.
Equation \ref{2terms} can also be derived from the replica trick \cite{Gromov:2014kia,Donnelly:2014gva,Schnitzer:2016lrd}. 

We note that there is an alternative \emph{algebraic} approach to studying entanglement in gauge theory in which one associates entropy to the algebra of gauge-invariant local observables \cite{Casini:2013rba}. 
Here one has to be careful about the precise definition of the algebra, and different choices are possible.
Generically, the local algebra will have a center, and one distinguishes the different algebras by their center.
For an abelian gauge theory, $S_\text{Boltzmann}$ vanishes and the entropy associated with the \emph{electric center} coincides with \eqref{2terms}.
However, for a nonabelian theory the algebraic entropy coincides with $S_\text{Shannon}$ and does not include the $S_\text{Boltzmann}$ term \cite{Soni:2015yga}.
Similarly, we can consider quantities such as the mutual information $I(\mathbf{A}:\mathbf{B}) = S(\mathbf{A}) + S(\mathbf{B}) - S(\mathbf{A} \cup \mathbf{B})$ between two intervals $\mathbf{A}$ and $\mathbf{B}$ separated by some finite distance.
The mutual information is sometimes viewed as a regularized version of the entanglement entropy \cite{Casini:2006ws}.
It is not hard to see that the Boltzmann entropy $S_\text{Boltzmann}$, which is exactly additive, cancels out of the mutual information leaving only the contribution from the Shannon entropy.
In this work we will consider both terms in \eqref{2terms} separately, and we will see that they behave quite differently in the large $N$ limit.

Several authors \cite{Donnelly:2016auv, Harlow:2016vwg, Lin:2017uzr} have noted the similarities between the expansion \eqref{2terms} and the holographic entanglement entropy formula \eqref{RT}.
Like the leading term in the Ryu-Takayanagi formula, the Boltzmann entropy term is both linear in the state and local to the entangling surface.
Moreover it is additive in the sense that it is linear in $m$, the number of points in the entangling surface.
Generically the entanglement entropy is neither linear nor local, which raises the question of why the leading order term of the large $N$ expansion has these properties. 

The goal of this paper is to understand the relation between the expansion of the entropy \eqref{2terms} and the asymptotic expansion at large $N$.
We therefore focus on the large $N$ behaviour of the different terms of the entropy. 
Our main question is: how do they each scale with $N$? 
Does one dominate over the other? 
It is a natural (and important) problem to consider, given that $S_{\text{Boltzmann}}$ arises purely from counting edge states; it is our analog of the Bekenstein-Hawking area term.
We hope that understanding the behaviour of different parts of entanglement entropy in the large $N$ limit of two-dimensional Yang-Mills will lead us to a better understanding of entropy in large $N$ gauge theories in general, and particularly those relevant for holography.

This paper is organized as follows. 
In section \ref{section:2dym}, we review two-dimensional Yang-Mills theory and the expression of the partition function as a sum over irreducible representations of the gauge group and specialize to the case where the gauge group is $\mathrm{U}(N)$.
We demonstrate how the entropy calculated via the replica trick splits into a sum of the local ``Boltzmann'' term and the nonlocal ``Shannon'' term as in \eqref{2terms}.

In section \ref{section:fermions}, we map the $U(N)$ Yang-Mills theory as a theory of $N$ interacting fermions on a one-dimensional lattice.
The fermions are subject to a confining quadratic potential, as well as a repulsive ``entropic force'' similar to eigenvalue repulsion appearing in random matrix theory.
In the fermion description the Shannon entropy term arises as the thermal entropy of the fermions, while the Boltzmann term is the expectation value of the ``entropic potential''.
The fermionic model thus gives a natural explanation for the large $N$ scaling of the different terms of the model: the Boltzmann term, being the expectation value of a pairwise interaction potential, is $O(N^2)$; while the the Shannon term, being the entropy of the fermions, is naturally $O(N)$.

In section \ref{section:n} we review the saddle point analysis of the large $N$ limit, following Douglas and Kazakov \cite{Douglas:1993iia}.
In the weak coupling/low temperature phase the fermions behave like eigenvalues of a random matrix, and their density follows a Wigner semicircle distribution.
At a critical value of the coupling there is a third-order transition to a strong coupling/low temperature phase where the fermions form a Fermi sea.
We calculate the leading part of the entropy coming from the saddle point, find agreement with the calculation of Ref.~\cite{Gromov:2014kia} up to a constant, and confirm this result with numerical simulations.
In this saddle point approximation we find that only the Boltzmann term contributes to the entropy; this shows that in this model, the analog of the ``area operator'' is the local operator which counts the logarithm of the number of edge states.

Section \ref{section:corrections} is dedicated to studying $1/N$ corrections to the entanglement entropy.
Recall that the $1/N$ expansion of the partition function takes a particularly simple form: in the weak coupling phase there are no perturbative $1/N$ corrections to the partition function \cite{Gross:1994mr}, while in the strong coupling phase the $1/N$ expansion of the partition function is an expansion in even powers of $1/N$ \cite{Gross:1992tu}. 
In fact this property of the large $N$ expansion was the first evidence for a string description of the theory further elucidated in Refs.~\cite{Gross:1993hu,Gross:1993yt,Taylor:1994zm}.
Surprisingly, we find that the Shannon entropy term scales \emph{linearly} with $N$.
While this would seem to be in conflict with the string expansion, we argue that it is not.
The odd powers of $N$ arise from taking the large $N$ limit \emph{after} the analytic continuation required by the replica trick.
In the special case that the replica trick does not modify the topology of the underlying surface, we find that the string expansion is saved by a precise cancellation between the order $N$ term of the Shannon entropy and the subleading order $N$ term in the Boltzmann entropy. 

We conclude in section \ref{section:discussion} with a discussion and summary of our results.

\section{Two-dimensional Yang-Mills theory}
\label{section:2dym}

Two dimensional Yang-Mills theory has no local degrees of freedom and can be solved exactly \cite{Migdal:1975zg,Witten:1992xu}.
It is therefore surprising that the theory is actually quite rich, particularly when put on the sphere.
At large $N$ and strong coupling, the theory can be described as a two-dimensional string theory in which the partition function can be expressed as a sum over branched covers of spacetime by a two-dimensional worldsheet \cite{Gross:1992tu,Gross:1993hu,Gross:1993yt}.
At a finite value of the coupling the string description breaks down, and the theory exhibits a third-order transition into a phase which can be understood in terms of a random matrix model \cite{Douglas:1993iia,Gross:1994mr}.
For details we refer the reader to the extensive review \cite{Cordes:1994fc}.

For our purposes, the main feature of Yang-Mills is that, given its almost topological status, the entanglement entropy is finite and can be computed explicitly with relative ease; to wit, it is the simplest gauge theory in which one can begin to deeply understand entanglement for general gauge theories. 

We consider two-dimensional Yang-Mills theory on a compact, orientable 2D Riemannian manifold.
To specify the theory we specify the gauge group $G$, which we will take to be $U(N)$, and the Yang-Mills coupling constant $g$.
We will be interested in the large $N$ limit, for which we introduce the 't Hooft coupling $\lambda = g^2 N$ which is held fixed as $N \to \infty$. 
On a manifold of Euler characteristic $\chi$ and total area $A$, the partition function is given by
\begin{equation} \label{Z}
Z(A, \chi) = \sum_{R} (\dim (R))^\chi e^{-\frac{\lambda A}{2 N} C_2(R)}.
\end{equation}
This depends only on the Euler characteristic and the total area, which is a large simplification.

To calculate the partition function explicitly using \eqref{Z}, we need a parametrization of the irreducible representations $R$ as well as their dimensions $\dim(R)$ and the quadratic Casimir $C_2(R)$.
Every irreducible representation of $\SU(N)$ has an associated Young diagram, usually represented as a collection of boxes arranged in $N$ rows of length $n_1 \geq n_2 \geq \cdots \geq n_N = 0$.
Each diagram corresponds to a tensor representation with $n$ indices, where $n = n_1 + n_2 + \ldots + n_N$ is the total number of boxes.
For example, the trivial representation corresponds to a diagram with no boxes, $n = 0$, while the fundamental representation is a tensor with one index and therefore corresponds to a diagram with one box, $n_1 = 1, n_2 = n_3 = \ldots = n_N = 0$.
A single row of length $k$ corresponds to a symmetric $k$-tensor, $n_1 = k, n_2 = n_3 = \ldots = n_N = 0$.
An antisymmetric $k$-tensor is a single column of height $k$, which is represented by a diagram with $n_1 = n_2 = \cdots = n_k = 1$ and $n_{k+1} = \cdots = n_N = 0$.
In $\SU(N)$ we require that $k < N$ because a column of height $N$ is proportional to $\det(U)$, which is trivial in $\SU(N)$.

To instead use $\U(N)$, we consider all products of each representation by multiples of $\det(U)$.
This amounts to shifting all the row lengths by a fixed constant, or equivalently, relaxing the restriction $n_N = 0$ and allow for an arbitrary sequence of row lengths $n_1 \geq \cdots \geq n_N \in \mathbb{Z}$.

In terms of the Young diagram, the expressions for $C_2(R)$ and $\dim(R)$ are (see e.g. \cite{Douglas:1993iia})
\begin{align}
C_2(R) &= \sum_{i = 1}^N n_i (n_i - 2i + N + 1), \label{C2R} \\
\dim(R) &= \prod_{1 \leq i < j \leq N} \left(1 - \frac{n_i - n_j}{i - j} \right). \label{dimR}
\end{align}
The distinction between $\SU(N)$ and $\U(N)$ only appears at order $N^0$ in the partition function, while we will primarily be interested in effects at $O(N^2)$ and $O(N)$.

\subsection{Entanglement entropy via the replica trick}

To calculate an entanglement entropy in the theory we will proceed by the replica trick.
Suppose we start with a 2D Riemannian manifold $M$ of Euler characteristic $\chi$ and area $A$.
We can cut it along a one-dimensional surface $\Sigma$ without boundary and we interpret the Euclidean path integral over $M$ as producing a mixed state on the surface $\Sigma$. 
In general, $\Sigma$ is the union of some number of circles.
We further divide $\Sigma$ into two disjoint parts $\Sigma = \mathbf{A} \cup \mathbf{B}$. 
Each of $\mathbf{A}$, $\mathbf{B}$ consists of some number of circles and some number of intervals. 
Let $m$ be the total number of points where we cut $\Sigma$ into two intervals; we call this collection of points the entangling surface.
We will be interested in the amount of entanglement between the states of the regions $\mathbf{A}$ and $\mathbf{B}$ in the state produced by the path integral over $M$.

Using the replica trick we form a new manifold $M_n$ by taking $n$ copies of $M$, cutting them along $\Sigma$, and regluing.
Let us denote the $n$ copies of $M$ by $M^i$, $i = 1, \ldots n$.
When we cut along $\mathbf{A}$, we introduce two boundaries, one on either side of the cut; we denote the two boundaries on $M^i$ by $\mathbf{A}_\pm^i$.
Similarly, by cutting along $\mathbf{B}$ we introduce boundaries $\mathbf{B}_\pm^i$.
To obtain the replicated manifold, we glue $\mathbf{B}_+^i$ to $\mathbf{B}_-^i$, and $\mathbf{A}_-^i$ to $\mathbf{A}_+^{i+1}$.
We continue the gluing cyclicly so that $\mathbf{A}_-^n$ gets glued back to $\mathbf{A}_+^1$. 
The resulting replicated manifold will have area $A_n$ and Euler characteristic $\chi_n$ where
\begin{equation} \label{Achi}
    A_n = n A, \quad \chi_n = n \chi + (1-n) m.
\end{equation}
The latter equation is not immediately obvious but follows from the inclusion-exclusion formula for the Euler characteristic.
We choose a triangulation of $M$ and apply Euler's formula $\chi = V - E + F$. 
Each edge and face of $M$ appears $n$ times in $M_n$, except for the points lying on the entangling surface which appear just once.
Correcting for this overcounting leads to the formula for $\chi_n$ \eqref{Achi}.

The (unnormalized) reduced density matrix of region $\mathbf{A}$ then satisfies
\begin{equation}
    \tr (\rho_\mathbf{A}^n) = Z(A_n, \chi_n)
\end{equation}
We find the entanglement entropy by differentiating with respect to the replica number $n$:
\begin{align}
    S &= \left( 1 - n \frac{d}{dn} \right) \log Z(A_n, \chi_n) \Big|_{n=1} \\
    &= \left( 1 - (\chi - m) \partial_\chi - A \partial_A \right) \log Z(A, \chi). \label{totalentropy}
\end{align}
One might worry that this does not make a lot of sense: we are taking derivatives with respect to the integer parameters $n$ and $\chi$.
To perform this differentiation, we have to analytically continue the partition function $Z$ to complex values of $\chi$.
In general such an analytic continuation from the integers would not be unique, but Carlson's theorem gives a set of sufficient conditions for uniqueness.
The present case is simpler: the sum \eqref{Z} is well-defined for complex $A$ and $\chi$ as long as the real part of $A$ is positive, so it can be taken as the definition of the analytic continuation.

To relate the entropy \eqref{totalentropy} to the fluctuations of the observable $R$ we introduce the probability distribution over representations
\begin{equation}
    p_R = \frac{1}{Z} \dim(R)^\chi e^{-\frac{\lambda A}{2N} C_2(R)}.
\end{equation}
This distribution determines what a local observer would detect by measuring gauge-invariant local operators.
In terms of the probability distribution $p_R$, the entropy takes the form
\begin{equation}
    S = S_\text{Shannon} + S_\text{Boltzmann},
\end{equation}
where the Shannon term is the entropy associated with fluctuations of the representation $R$,
\begin{align}
    S_\text{Shannon} &= (1 - \chi \partial_\chi - A \partial_A) \log Z(A,\chi) \label{shannon}
    \\ &= - \sum_{R} p_R \log p_R,
\end{align}
while the Boltzmann term is associated with the indistinguishable microstates residing at the endpoints of the intervals,
\begin{align}
    S_\text{Boltzmann} &= m \, \partial_\chi \log Z(A, \chi) \\
    &= m \sum_{R} p_R \log \dim(R).
\end{align}
Note that this term is local in the sense that it can be written as the expectation value of the local operator $\log \dim(R)$ summed over the $m$ points of the entangling surface.

An important special case is the two-dimensional de Sitter entropy, which corresponds to an entangling surface that is a single interval on a sphere for which $\chi = m = 2$ \cite{Donnelly:2014gva}.
Another special case is the thermal entropy of 2D Yang-Mills theory on a spatial circle, for which $\chi = m = 0$.
Some more general cases were considered in \cite{Donnelly:2018ppr}.

\section{Fermion description of two-dimensional Yang-Mills} \label{section:fermions}

To study this concretely, it is useful to map the problem to a system of fermions.
The fermion mapping was studied in \cite{Minahan:1993np,Douglas:1993xv,Douglas:1993wy} for Yang-Mills theory on a torus, i.e. at $\chi = 0$.
Here we will allow for $\chi > 0$.

Let us consider a system of $N$ fermions which can occupy sites on a 1D lattice labelled by integers.
Since fermions are indistinguishable and cannot occupy the same site, we can assign each configuration a sequence $h_1, \ldots, h_N$ of fermion positions where $h_1 < h_2 < \cdots < h_N$ (i.e. the Pauli exclusion principle is satisfied).\footnote{When studying the model on a torus, the label $h_i$ was identified with the momentum of a nonrelativistic free fermion.
When $\chi \neq 0$ there is an additional potential term which is nonlocal in $h$, and it is more natural to think of this as a potential in position space rather than in momentum space.
}
We can associate each fermion configuration to a Young diagram by defining
\begin{equation}
    n_i = i - h_i + c,
\end{equation}
where $c$ is a constant that we are free to choose. 
Under this mapping the difference between successive row lengths of the Young tableau is the number of empty spaces between fermions.

In terms of the fermion positions $h_i$, the quadratic Casimir and dimension of the representation (\ref{C2R}, \ref{dimR}) take a simpler form:
\begin{align}
C_2(R) &= \sum_{i} h_i^2 -\frac{1}{12}N(N^2-1) \label{C2}  \\ 
\log \dim(R) &= \sum_{i < j} \log \left(h_j - h_i \right) -\sum_{i<j} \log(j-i). \label{logdim}
\end{align}
Here we have chosen $c=-(N+1)/2$ to eliminate a linear term in the potential \eqref{C2}.
The constant terms ensure that the trivial representation has $C_2(R) = 0$ and $\log \dim(R) = 0$.
The constant term in $C_2$ is just shifts the ground state energy, and does not change the entropy or any other physical observable. 
However, the constant appearing in $\log \dim(R)$ is important as it contributes directly to the entropy.

In terms of the fermion system, we can write the partition function as
\begin{equation} \label{Zfermion}
    Z_\text{fermion}(\beta) = \sum_{h_1 < h_2 < \cdots < h_N} \exp(- \beta E(h_1, \ldots, h_N)),
\end{equation}
where $\beta$ is the inverse temperature of the fermion system. 
Given that $h$ labels the sites in a one-dimensional lattice, the $C_2$ term in the energy represents a confining external quadratic potential while the $\log \dim(R)$ term is a repulsive potential between the fermions akin to eigenvalue repulsion in random matrix theory.

Physically, this system corresponds to a wire composed of a discrete number of sites in two spatial dimensions.
The electrons repel each other by the Coulomb potential (which is logarithmic in 2D), and are confined in a quadratic potential $\phi \sim r^2$ (which is what we would get at fixed density $\rho$).
Therefore the potential energy of a configuration is the sum of the external potential and the interaction potential
\begin{equation}
E = E_\text{ext} + E_\text{int} = \sum_{i} \frac{1}{2} q \rho a^2 h_i^2 - \sum_{i < j} \frac{q^2}{2 \pi} \log\frac{a(h_j - h_i)}{C}.
\end{equation}
Here $q$ is the charge of the fermions, $a$ is the lattice spacing, and $C$ is a constant with dimensions of length.
This gives a mapping between the variables of the fermionic model ($q,\rho,a$ and $\beta$) and the Yang-Mills theory ($A$, $\lambda$ and $\chi$).
This mapping is not one-to-one because the parameters have units, but we can identify dimensionless combinations of parameters of the two models.

In this simple picture we can already intuitively see why the system has two phases.
The weak coupling phase occurs when the logarithmic repulsion term dominates over the external potential creating a low fermion density.
As the coupling constant is increased, the fermions are compressed together and eventually the density in the center reaches a maximum because of the Pauli exclusion principle.
This creates a phase transition, as the strong coupling forces the electrons into a ``Fermi sea'' in the middle of the wire.

We can now study the entanglement entropy of the gauge theory in the language of this fermion model.
To obtain the thermal entropy of the fermion system we vary $\beta$ in the partition function \eqref{Z}.
In the Yang-Mills theory this corresponds to a simultaneous variation of $A$ and $\chi$:
\begin{align}
    S_{\text{fermion}} &\equiv (1-\beta \partial_{\beta})\log Z_\text{fermion}(\beta) \\
    &= (1 - A \partial_A - \chi \partial_\chi) Z(A,\chi) \\
    &= S_{\text{Shannon}}.
\end{align}
Thus we find the fluctuations of the representations in Yang-Mills are captured by the thermal fluctuations in the Fermi gas. 
Furthermore, in the fermion model,
\begin{align} \label{Sboltzfermion}
    S_{\text{Boltzmann}} = m\braket{\log\dim(R)} \propto \langle E_\text{int} \rangle,
\end{align}
i.e., the Boltzmann entropy of the Yang-Mills theory is proportional to the fermion interaction energy.

The discrepancy between the entropy of the fermion model and the entropy of Yang-Mills theory originates in the treatment of the $\log \dim(R)$ term.  In the fermion model we treat this as a force term, so it does not contribute to the entropy.
The expectation value of the ``entropic potential'' that gives rise to this entropic force is precisely the missing entropy \eqref{Sboltzfermion}.

The fermion model gives some insight into the large $N$ scaling of the two terms in the entropy.
The Shannon entropy is the entropy of a system of $N$ fermions, so we expect it to scale linearly with $N$.
The Boltzmann entropy term from the fermion perspective is the expectation value of the interaction energy; since there is a contribution for each pair of fermions we expect that this term scales as $N^2$.
This intuitive picture will be borne out by the precise calculations in the following sections.

\section{Large $N$ limit}
\label{section:n}

In this section we analyze the entropy of two-dimensional Yang-Mills at leading order in the large $N$ expansion.
We find the saddle point configuration and comment on the phase transition found by Douglas and Kazakov \cite{Douglas:1993iia}, correcting the value of the constant term in the free energy.
We find the entropy to leading order in the large $N$ limit, which agrees with the result of \cite{Gromov:2014kia} up to the aforementioned constant.
To conclude the section we give a direct argument showing that only the Boltzmann entropy contributes at leading order, while the Shannon entropy is subleading.
We find it is a consequence of the fact that the partition function is dominated by a single saddle point.

\subsection{Continuum limit}

In taking the continuum limit, it will be useful to start from the fermionic description, for which the partition function is given by
\begin{align}
    Z &=\sum_{\{h_i\}}\exp\left\{-\frac{\lambda A}{2N} \left(\sum_{i=1}^N h_i^2\right) + \chi \left(\sum_{i<j}^N  \log(h_j-h_i)\right) + \frac{\lambda A}{24}(N^2-1)-\chi \sum_{i<j} \log (j-i)\right\}.
\end{align}
The last two terms are independent of $h$ and simply shift the ground state energy and entropy.

Now we change variables to $x_i\equiv i/N$, $h(x_i)\equiv h_i/N$ and take $N \to \infty$. We also define a continuum density of fermions $\rho(h) = \frac{\partial x}{\partial h}$ which is bounded between zero and one.
In the large $N$ limit the sums appearing in the partition function become integrals, with the outer sum over $\{h_i\}$ turning into a path integral over all functions $h(x)$ satisfying $h(x)-h(y)\geq x-y$ (which just enforces the fact that the density of fermions is bounded by 1).
The continuum partition function then takes the form
\begin{align}
    Z = \int \mathcal{D}[h] \exp(-N^2 S_{\text{eff}}[h]). \label{path}
\end{align}
The measure $\mathcal{D}[h]$ in this path integral is determined because it arises from a discrete sum for which the measure is fixed.
The action associated with a fermion configuration is
\begin{align}
    S_{\text{eff}}[h] = \frac{\lambda A}{2} \int_0^1  dx h^2(x)  - \chi \int_{2\epsilon}^1 dy  \int_\epsilon^{y-\epsilon}dx \log\left|\frac{h(y)-h(x)}{y-x}\right| -\frac{\lambda A}{24}. \label{Seff}
\end{align}
The parameter $\epsilon$ comes into the expression because the double sum was over $i < j$, leading to a cutoff $|x-y| > 1/N$. 
Taking $\epsilon\to 0^+$ creates a principal value integral
\begin{align}
S_{\text{eff}}[h] = \frac{\lambda A}{2} \int_0^1  dx h^2(x)  -\frac{\chi}{2}\int_0^1 dy \fint_0^1 dx \log\left|\frac{h(y)-h(x)}{y-x}\right| -\frac{\lambda A}{24}.
\end{align}

Since the exponential term in \eqref{path} scales as $N^2$, the method of steepest descent gives a good approximation to the partition function for large $N$.  This means we should find a minimum of the effective action.
Consider a variation $h(z)\to h(z)+\delta h(z)$ and keep only first order terms:
\begin{align}
    \delta S_{\text{eff}} = \lambda A\int_0^1 dx \delta h(x) h(x) - \chi\int_0^1 dx \delta h(x) \fint_0^1 \frac{dy}{h(x)-h(y)}.
\end{align}
This leads to the saddle point equation
\begin{align}
    \frac{\lambda A}{\chi} h = \fint_{-a}^a \frac{\rho(s) ds}{h-s},
    \label{eq:integralequation}
\end{align}
where we have made a change of variables $dy=\rho(s)ds$ to work with the density rather than the fermion positions, and we integrate from $-a$ to $a$, assuming on physical grounds that $\rho$ vanishes outside of some finite interval $(-a,a)$.
This equation is simply the condition that the force coming from the external potential balances against the entropic force coming from the other fermions.
Notice that in the interval of interest, $h \in [-a,a]$, the integrand has a pole.

In addition to the saddle point equation \eqref{eq:integralequation}, we will have to impose boundary conditions to arrive at the solution. 
We also need to ensure that the Pauli exclusion principle is satisfied, i.e. $\rho(s)\leq 1$. 
This constraint is ultimately what provokes a phase transition.

\subsection{Ground state configuration}

We are interested in solving the saddle point equation, which will tell us the ground state configuration of our system.
This was done by Douglas and Kazakov \cite{Douglas:1993iia}, who solved the theory at large $N$ and found that it has a phase transition.
Note that if we restrict to physical values of $\chi = 2 - 2g$, the phase transition is only present on the sphere, i.e. at $\chi=2$. 
However, for later convenience we will allow $\chi$ to take arbitrary positive values.

We begin by finding the ground state for the weak coupling phase, assuming that $\rho(s)$ is an analytic function and defining the resolvent
\begin{align}
    R(z) = \int_{-a}^a ds \frac{\rho(s)}{z-s} \label{R}.
\end{align}
The Sokhotski–Plemelj theorem relates the value of the resolvent at the branch cut to the density $\rho$ and is given by
\begin{align}
R_{\pm}(h) = \fint_{-a}^a ds \frac{\rho(s)}{h-s} \mp i \pi \rho(h) ,
\end{align}
where $R_{\pm}(h) = \lim_{\epsilon\to 0} R(h\pm i\epsilon)$. 
So our task is to find $R_{\pm}$, which will immediately give us $\rho$.

From the Cauchy integral formula we can write
\begin{align}
    R(z) = \oint_{C_z}\frac{dw}{2\pi i} \frac{R(w)}{w-z}\frac{g(z)}{g(w)} \label{cauchy}
\end{align}
where we choose $g(z)=z\sqrt{1-a^2/z^2}$ because it has the same branch cut as the integrand in (\ref{eq:integralequation}). Since $g$ has a branch cut for $z\in[-a,a]$, the contour $C_z$ (which surrounds $z$) must exclude this interval. 
After some algebra and contour manipulation (which can be found in the appendix), we arrive at
\begin{align}
    \rho (h) = -\lim_{\epsilon \rightarrow 0} \frac{R(h + i \epsilon) - R(h - i \epsilon)}{ 2\pi i} = \frac{\lambda A}{\pi \chi}\sqrt{a^2 - h^2},\label{rho_weak}
\end{align}
the famous Wigner semicircle.
Now to solve for the parameter $a$ we impose the constraint \begin{align}
    \int_{-\infty}^\infty dh \rho(h) = 1,
\end{align}
which gives $a=\sqrt{2\chi/\lambda A}$. 

For the strong coupling phase, we no longer assume $\rho(h)$ is analytic in all of its support. This is because it can hit a ``roof'' at $\rho=1$, which may create a discontinuity. Indeed, the solution we found earlier only satisfies the $\rho\leq 1$ constraint for $2 \lambda A \leq \chi \pi^2$ (which is why we call it the weak coupling phase). 
If this inequality is not satisfied, the Wigner semicircle is no longer a physical solution.

Instead we can look for a solution in which the distribution saturates the $\rho(h)=1$ bound near the origin and make the following ansatz:
\begin{align}
    \rho(h) = \begin{cases}
    1 & |h| < b, \\
    \tilde{\rho}(h) & b<|h|<a, \\
    0 & |h|>a,
    \end{cases}
\end{align}
with $\tilde{\rho}(h)$ analytic. Inserting this into (\ref{eq:integralequation}) gives
\begin{align}
    \frac{\lambda A}{\chi} h +\log\frac{h-b}{h+b} = \fint \frac{\tilde{\rho}(s)  ds}{h-s},\label{strong}
\end{align}
where the principal value integral now runs over $[-a,-b]$ and $[b,a]$. 
We define the resolvent,
\begin{align}
    \tilde{R}(z) = \int_{(-a,-b)\cup (b,a)} dh \frac{\tilde{\rho} (h)}{z - h} = \oint_{C_z} \frac{dw}{2 \pi i} \frac{\tilde{R}(w)}{w-z} \frac{g(z)}{g(w)} \label{Rtilde}
\end{align}
using the Cauchy integral formula as before, and choose the function 
\begin{align}
    g(z) = z^2 \sqrt{1 - \frac{a^2}{z^2}} \sqrt{1 - \frac{b^2}{z^2}} \label{gz}
\end{align}
which has branch cuts at $[-a,-b]$ and $[b,a]$. After some algebra and contour integration which can again be found in the appendix, we arrive at
\begin{equation} \label{rho_strong}
    \rho(h) = \begin{cases} 1 & |h| < b \\
    \frac{2}{\pi a |h|} \sqrt{a^2- h^2} \sqrt{h^2 - b^2} \, \Pi\left( \frac{b^2}{h^2} \middle| \frac{b^2}{a^2} \right) & b < |h| <  a \\
    0 & |h| > a
    \end{cases}
\end{equation}
where $\Pi(n|m)$  is the complete elliptic integral of the third kind \eqref{ellipticPi}.
The constants $a$ and $b$ can now be determined in terms of $\alpha = \lambda A / \chi$ by applying two conditions.
The first condition is that there is no term linear in $z$ in the resolvent,
\begin{equation} \label{condition1}
    \alpha = \frac{2}{a} K \left( \frac{b^2}{a^2} \right),
\end{equation}
and the second is the normalization condition,
\begin{equation} \label{condition2}
\int_{-\infty}^\infty dh \rho(h) = a \left(2 E\left(\tfrac{b^2}{a^2}\right) + \left(\tfrac{b^2}{a^2}-1\right) K\left(\tfrac{b^2}{a^2}\right)\right) = 1.
\end{equation}
Here $K(m)$ and $E(m)$ are the complete elliptic integrals of the first \eqref{ellipticK} and second \eqref{ellipticE} kinds.

Given $\alpha$, these equations determine $a$ and $b$ as follows.
We first eliminate $a$ by multiplying \eqref{condition1} and \eqref{condition2} together, yielding a single equation to be solved for $m = b^2/a^2$.
While this equation does not appear to admit an analytic solution, it is monotonic and can be easily solved numerically by root finding.
The value of $a$ can then be found directly from \eqref{condition1}.

\begin{figure}[ht!]
    \centering
    \includegraphics[width=4in]{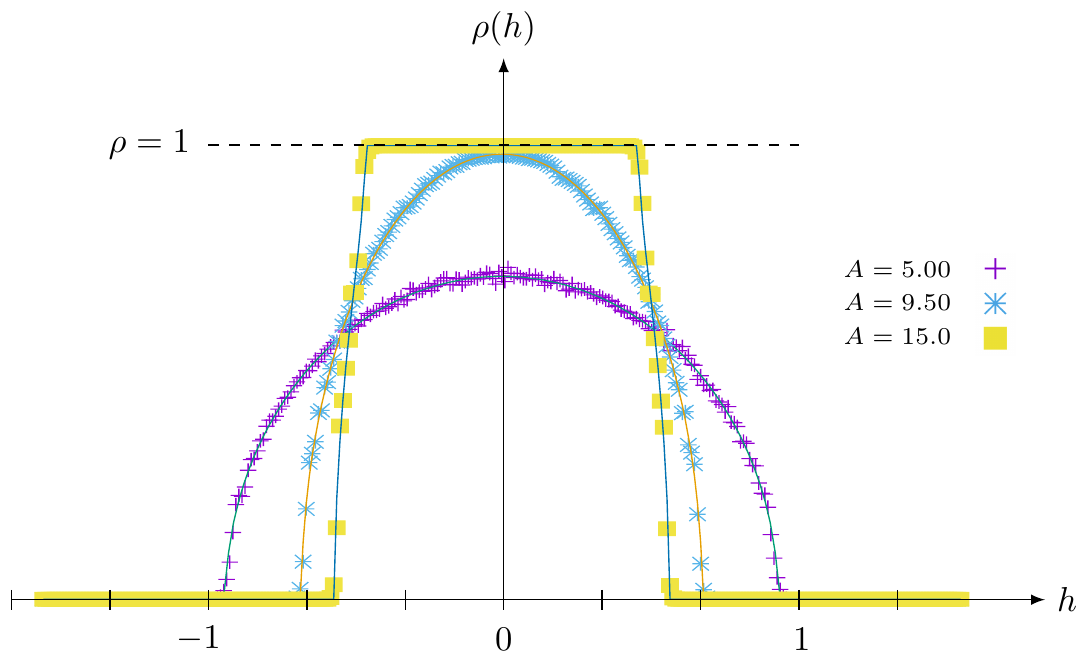}
    \caption{Different ground state configurations. The solid lines are the analytical functions, while the points represent data gathered from Monte Carlo simulations.
    Here we have set $\chi = 2$ and $\lambda = 1$, so $A$ is dimensionless.}
    \label{gnd_state}
\end{figure}

\subsection{Free energy and entropy}
\label{section:free-energy}

With the ground state configuration found, we can now make the saddle point approximation
\begin{align}
    Z\simeq \exp(-N^2 S_{\text{eff}}[h_0]),
\end{align}
where $h_0(x)$ is the configuration resulting from the ground state density $\rho(h)$ found in the previous section, and the effective action is given by
\begin{align}
S_\text{eff}[h_0]
    &= \frac{\lambda A}{2}\int_{-a}^a dh \rho(h)h^2 -\frac{\chi}{2} \int_{-a}^a dh \rho(h) \fint_{-a}^a dh' \rho(h') \log|h-h'| -\frac{\lambda A}{24} -\frac{3\chi}{4}.  \label{first_F}
\end{align}
Since we are working with the ground state density $\rho(h)$, we can use the saddle point equation
\begin{align}
    \frac{\lambda A}{\chi}\eta = \fint_{-a}^a ds \frac{\rho(s)}{\eta-s},
\end{align}
and integrate it with respect to $\eta$ from 0 to $h$,
\begin{align}
    \frac{\lambda A}{2\chi}h^2 = \fint_{-a}^a ds \rho(s)(\log|h-s| - \log |s|   ).
\end{align}
Plugging into (\ref{first_F}), we get
\begin{align}
    S_\text{eff}[h_0] = \frac{\lambda A}{4} \int_{-a}^a dh \rho(h) h^2 - \frac{\chi}{2} \fint_{-a}^a dh\rho(h) \log|h| - \frac{\lambda A}{24} -\frac{3\chi}{4}
\end{align}

\begin{figure}[ht!]
    \centering
    \includegraphics[width=4in]{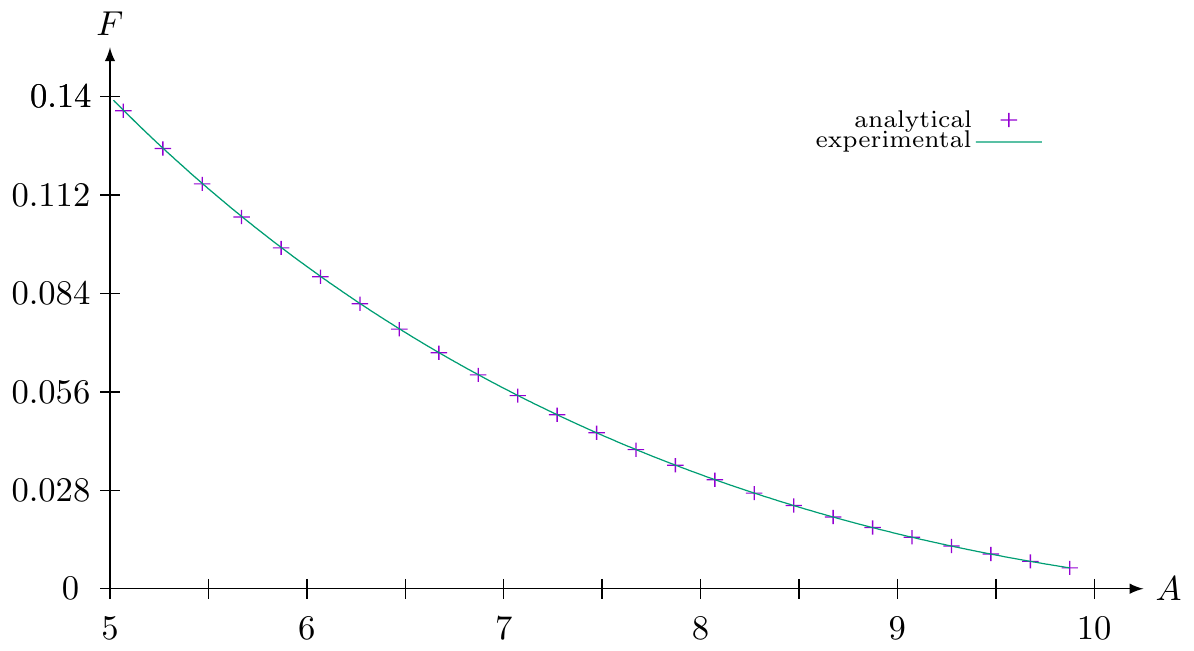}
    \caption{The free energy in the weak-coupling phase calculated from (\ref{final_F}) and from the Monte Carlo simulation as functions of $A$ at $N=31$ with $\lambda=1$ and $\chi = 2$.}
    \label{graph_F}
\end{figure}

In the weak coupling phase, we can use the solution \eqref{rho_weak} for $\rho(h)$ to find
\begin{align}
    \log Z(A, \chi) = N^2\left(\frac{3 \chi}{8} - \frac{\chi}{4} \log\left (\frac{2 \lambda A}{\chi}\right) + \frac{\lambda A}{24} \right)
    \label{final_F}
\end{align}
to leading order in $N$.
We note that this result is in disagreement with the one found in \cite{Gromov:2014kia}; however, we find agreement with a numerical evaluation of the partition function, as shown in figure \ref{graph_F}. 
The saddle point approximation for the free energy in the strong coupling regime does not yield an expression in terms of elementary functions, but it can be evaluated by numerical integration and also shows agreement with the simulation. 

Now we can readily show that to leading order in $N$, the only term that contributes to the entropy is indeed the Boltzmann term. 
We do not need the precise functional form of the partition function $Z$; it suffices to realize that the parameters of the problem enter the saddle point in the ratio $\lambda A/\chi$, so the saddle point $\rho(h)$ depends on the parameters of the problem through this ratio.
This is manifest in the formula \eqref{final_F}, but holds in both the weak and strong coupling phases. 
This implies that in the saddle point approximation the partition function is of the form
\begin{align}
    Z\simeq \exp\left(N^2(\lambda A f(\lambda A/\chi) +\chi g(\lambda A/\chi)\right),
\end{align}
where $f$ and $g$ are some functions obtained by integrating $\rho$ which only depend on the ratio $\lambda A/\chi$. 
It is then easy to show that for $Z$ of this form
\begin{align}
    S_{\text{Shannon}} = (1-\chi\partial_{\chi} -A\partial_A) \log Z = 0.
\end{align}

\begin{figure}[ht!]
    \centering
    \includegraphics[width=4in]{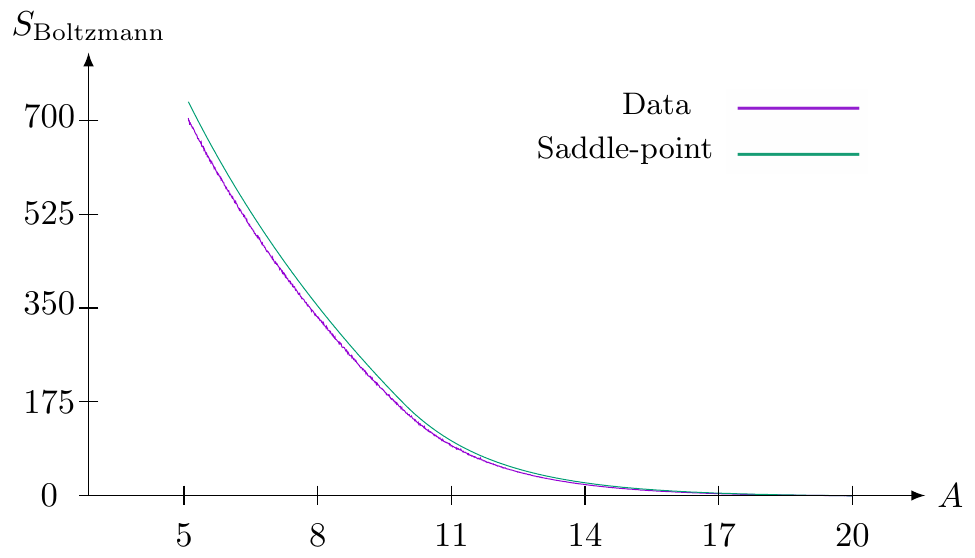}
    \caption{The Boltzmann entropy from a numerical simulation at $N = 41$ is compared with the prediction from the large $N$ saddle point.
    We see that the simulated Boltzmann entropy lies slightly below the saddle point value.
    This is consistent with the fact that the first subleading correction in the $1/N$ expansion is negative, as we will see in section \ref{section:corrections}. Here we have set $\chi = 2$ and $\lambda = 1$, so $A$ is dimensionless.
    }
    \label{figure:boltzmann}
\end{figure}

Then the Boltzmann term clearly dominates the entropy at leading order in $N$ for both strong and weak coupling. 
In the weak coupling phase we have the explicit formula
\begin{align}
    S_{\text{Boltzmann}} = m N^2 \left(\frac{1}{4} \log\frac{\chi}{2\lambda A} + \frac{5}{8} \right).
\end{align}
In the strong coupling phase we have no explicit formula, but we can still obtain saddle point results by a simple inversion; the result is plotted in figure \ref{figure:boltzmann}.
The key result is that the leading term in the entanglement entropy for large $N$ comes solely from the local term counting the edge modes: it is linear in the density matrix and proportional to the number of entangling points $m$.

\section{$1/N$  corrections}
\label{section:corrections}
    
We now consider $1/N$ corrections to the entropy, beyond the leading order results of the previous section.
In particular, we will calculate both the leading term in $S_\text{Shannon}$ and subleading terms in the $1/N$ expansion of $S_\text{Boltzmann}$.
Our results were produced by numerical simulations which we will describe in section \ref{subsection:numerics}; in section \ref{subsection:approximate} we will give some analytical checks which show agreement with the simulations.

What should we expect from the subleading corrections?
In the weak coupling phase, Gross and Matytsin studied subleading corrections to the sphere partition function and found that the $1/N$ expansion consists of only a leading order $N^2$ term, a subleading $N^0$ term and nonperturbative corrections, the largest of which is $O(N^{-1/2} e^{-N})$ \cite{Gross:1994mr}.
The strong coupling case is more interesting: Gross and Taylor \cite{Gross:1993hu} showed that the large $N$ expansion of $Z$ in the strong coupling phase could be organized as the partition function of a closed string theory. 
The genus expansion of the partition function is an expansion in even powers of $N$, where the term of order $N^\chi(\Sigma)$ counts branched covers of the target space by a worldsheet $\Sigma$.
In both phases, the perturbative $1/N$ expansion contains only even powers of $N$.

However, surprisingly, the Shannon entropy and the Boltzmann entropy each have a term linear in $N$.
We argue that this is a consequence of the analytic continuation required by the replica trick.
The replica trick can change the topology of spacetime, and the powers of $N$ appearing in the partition function therefore depend on the replica number.
Thus there is an issue with the order of limits; the correct thing is to do the replica trick at finite $N$ and only then do the asymptotic expansion at large $N$.

\subsection{Numerical method}
\label{subsection:numerics}

The main approach we used is the Markov Chain Monte Carlo (MCMC) method, see e.g. \cite{Hanada:2018fnp} for a review.
The state space of our system consists of a list of fermion positions $h_1 > \cdots > h_N$.
Using the definition of $C_2(R)$ and $\log \dim(R)$ in terms of fermion positions, we compute the Boltzmann factor $\exp(-\beta E_1)$ for the configuration. 
A new candidate fermion configuration is proposed by randomly shifting the position of one of the fermions by one lattice site.
We then compute the new energy $E_2$, and switch to the new configuration if the quantity $\exp(-E_2+E_1)$ is greater than a random number between 0 and 1.\footnote{Accompanying the Boltzmann factor is also a factor related to the number of empty spots adjacent to fermions (``edges'') in each configuration.
}
The detailed balance condition then ensures that after an initial ``burn-in'' period, the distribution of configurations follows the Boltzmann distribution.
To reduce the burn-in time, we initialize the fermion configuration so that the density of fermions approximates the analytic form \eqref{rho_weak}, \eqref{rho_strong} derived in the large $N$ limit.

The MCMC method allows us to sample from the probability distribution of fermion configurations, but does not allow us to find the partition function or entropy directly.
Instead we calculate derivatives of the partition function by using expectation values
\begin{align}
    \partial_A \log Z &= - \left \langle \frac{\lambda C_2(R)}{2 N} \right \rangle,\\
    \partial_\chi \log Z &= \langle \log \dim(R) \rangle.
\end{align}
Similarly, second derivatives of the partition function are encoded in the (co)variances
\begin{align}
\partial_A^2 \log Z &= \left \langle \left( \frac{\lambda C_2(R)}{2 N} \right)^2 \right \rangle_c ,\\
\partial_\chi^2 \log Z &= \left \langle ( \log \dim (R))^2 \right \rangle_c , \\
\partial_A \partial_\chi \log Z &= -\left \langle \frac{\lambda C_2(R)}{2N} \log \dim(R) \right \rangle_c .
\end{align}
where we define the connected correlation functions as $\langle \mathcal{O}_1 \mathcal{O}_2 \rangle_c = \langle \mathcal{O}_1 \mathcal{O}_2 \rangle - \langle \mathcal{O}_1 \rangle \langle \mathcal{O}_2 \rangle$.

This lets us find the Shannon entropy in the following way, which we denote the {\it variance method}. 
The derivatives of the Shannon entropy are given by
\begin{align}
\partial_A S_\text{Shannon} &= \left \langle \left(\frac{\lambda C_2(R)}{2N} \right) \left( -A \frac{\lambda C_2(R)}{2N} + \chi \log \dim(R) \right) \right \rangle_c,
\\
\partial_\chi S_\text{Shannon} &= -\left \langle (\log \dim(R)) \left( -A \frac{\lambda C_2(R)}{2N} + \chi \log \dim(R) \right) \right \rangle_c.
\end{align}
We can therefore carry out the simulation for various values of $A$ and integrate $\partial_A S_{\text{Shannon}}$ numerically to find the Shannon entropy.

Since we have two parameters, $A$ and $\chi$, we can integrate the Shannon entropy along any curve in this two-dimensional space of couplings.
This leads to a method we call the {\it first law method}.
Recall that varying $A$ and $\chi$ simultaneously does not change the saddle point configuration, which prompts the consideration of the combination
\begin{equation} \label{AdAchidchi}
(A \partial_A + \chi \partial_\chi) S_\text{Shannon} = - \left \langle \left(-A \frac{\lambda C_2(R)}{2 N} + \chi \log \dim(R) \right)^2 \right \rangle_c.
\end{equation}
To find $S_\text{Shannon}$ we can start from $(A,\chi)$, and integrate the derivative \eqref{AdAchidchi} along the line $( A/\tau, \chi/\tau)$.
More precisely, we define a function $f(\tau) = S_\text{Shannon}\left( \frac{A}{\tau}, \frac{\chi}{\tau} \right)$ which then satisfies
\begin{equation}
    f'(\tau) =  \frac{1}{\tau^2} \left \langle \left(-A \frac{\lambda C_2(R)}{2 N} + \chi \log \dim(R) \right)^2 \right \rangle_c
\end{equation}
where the variance is calculated at area $A/\tau$, and Euler characteristic $\chi/\tau$.
Integrating this from 0 to 1 allows us to find $S_{\text{Shannon}}$.
This method has the added benefit that starting at $\tau = 1$ and decreasing to $\tau = 0$ causes the simulation to converge to the equilibrium distribution more efficiently---a phenomenon known as annealing. 
We call this the first law method because we are essentially integrating the first law of thermodynamics for the fermion system: $S = \int_0^1 d \langle E \rangle/\tau$ where $E = \frac{A \lambda}{2N} C_2(R) - \chi \log \dim(R)$ is the energy of the fermion system. 

\begin{figure}[ht!]
    \centering
    \includegraphics[width=4in]{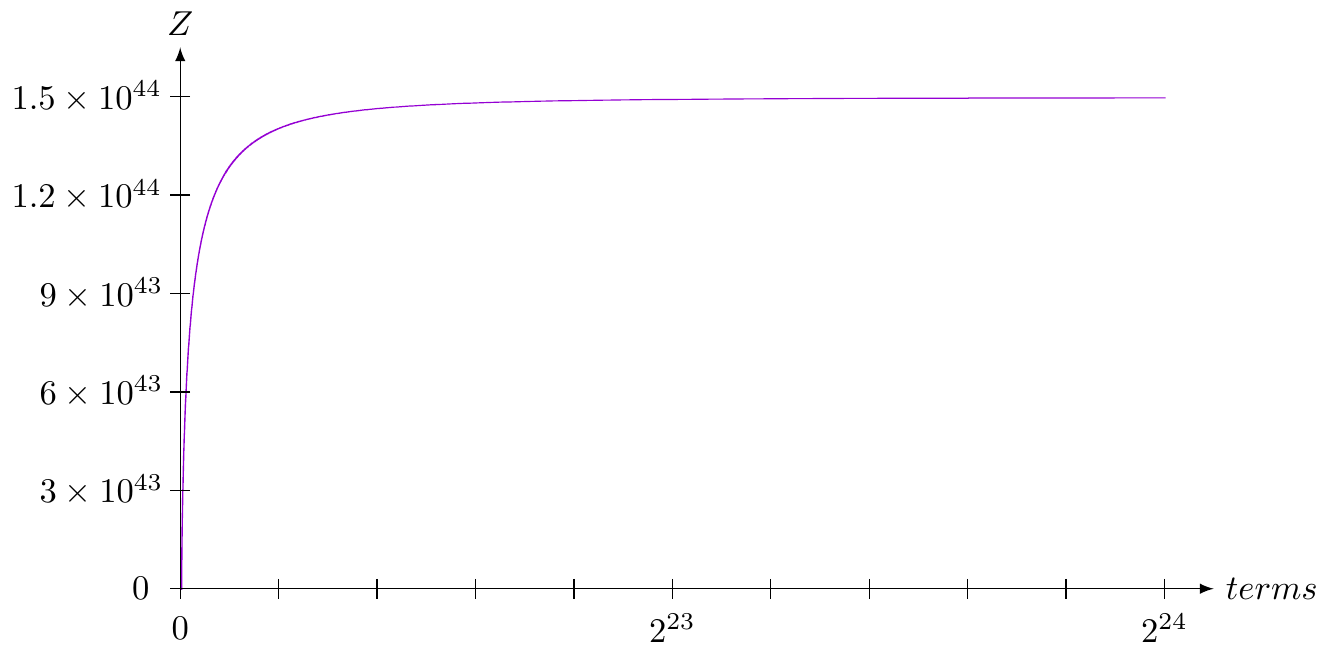}
    \caption{The partition function calculated from brute force addition as a function of the number of lowest energy terms included in the sum at $\lambda A=8.00$ for 21 fermions.}
    \label{figure:Z_brute}
\end{figure}

We implemented both the variance and first law methods, and found they agree.
However, we found the Monte Carlo method does not always converge at sufficiently strong coupling; we attribute this to the fact that at strong coupling the potential is very steep, increasing the probability that the system will get stuck in a local minimum.
This led us to consider a ``brute force addition'' method, where we priority queue a list of configurations, sorted by their Boltzmann weights. At each iteration we remove the configuration with the largest weight, mark it as visited, then add all unvisited neighbouring configurations to the queue.
This allows us to iterate over configurations in increasing energy order without repetition, from which we can directly calculate the partition function and entropy.
We truncate the sum once doubling the number of terms added changes the total by less than 1\%; this leads to good convergence, as illustrated in figure \ref{figure:Z_brute}.
At weak coupling, the brute force method converges more slowly, but in this regime the MCMC simulation is efficient.
At intermediate value of the coupling, where both methods are reliable, we also found agreement.

\subsection{Numerical results}
\label{subsection:numericalresults}

\begin{figure}[ht!]
    \centering
    \includegraphics[width=4in]{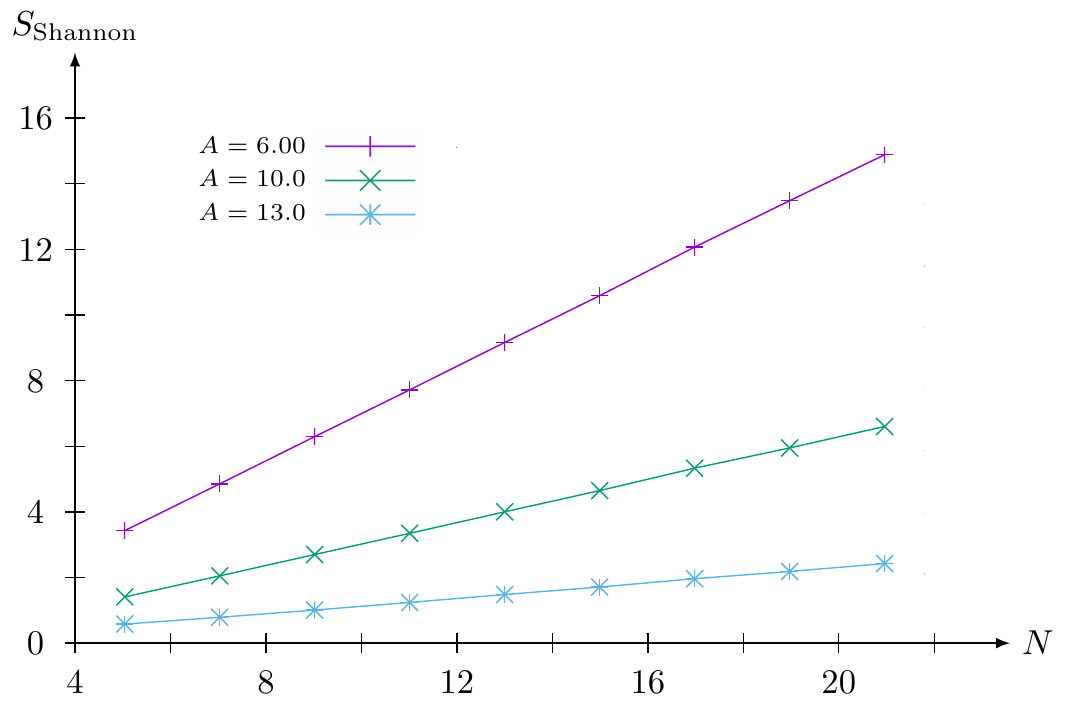}
    \caption{Shannon entropy as a function of $N$ obtained using the variance method shows a clear linear scaling with $N$.}
    \label{figure:shan_fixedA}
\end{figure}

\begin{figure}[ht!]
    \centering
    \includegraphics[trim={0 2.5cm 0 1.5cm},width=4in,clip]{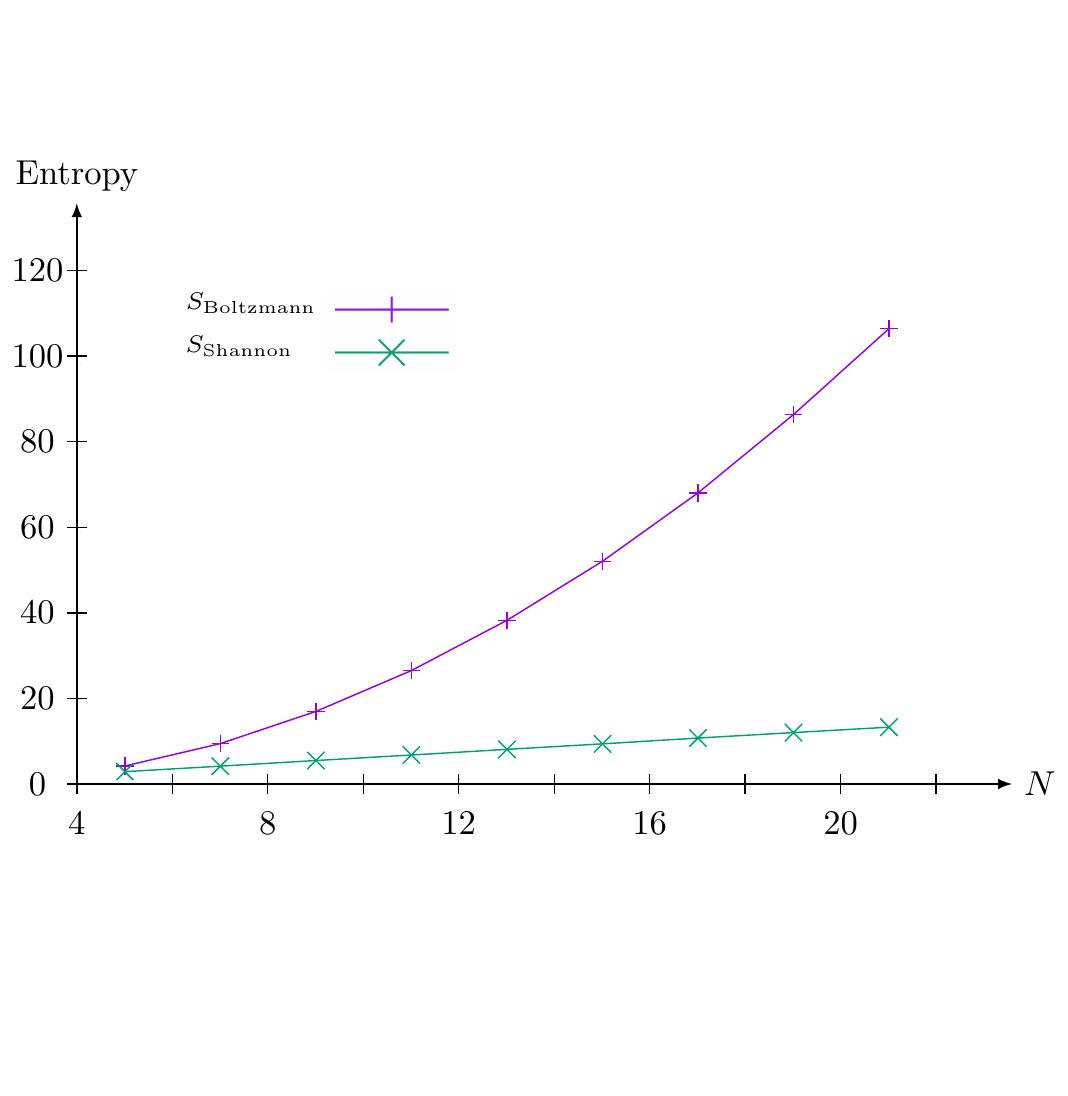}
    \caption{Boltzmann and Shannon entropies as functions of $N$ at $\lambda A=7$. 
    The linear scaling of the Shannon entropy is in clear contrast with the quadratic scaling of the Boltzmann entropy.
    \label{figure:shan_boltz}
    }
\end{figure}

\begin{figure}[ht!]
    \centering
    \includegraphics[width=4in]{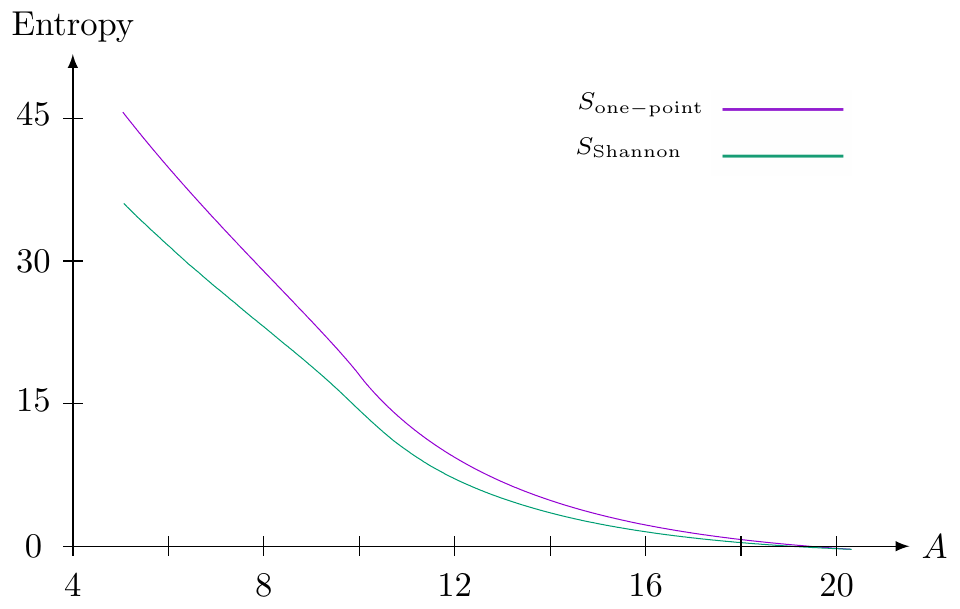}
    \caption{Shannon and one-point entropies as  functions of $A$ at $N=41$ (with $\lambda=1$). Note the phase transition at $A=\pi^2$. Reference point at $A=20$.}
    \label{figure:shan_fixedN}
\end{figure}

From the simulations of the previous section we can extract both the way that $S_{\text{Shannon}}$ scales with $N$ and its dependence on $A$.

In figure \ref{figure:shan_fixedA} we plot the Shannon entropy as a function of $N$ for several values of the coupling, and see a clear linear scaling with $N$.
This can be compared with the quadratic scaling of the Boltzmann entropy shown in figure \ref{figure:shan_boltz}.
The $N$ dependence of the Shannon entropy is surprising because it means $S_{Shannon} =(1-A\partial_A-\chi\partial_\chi)\log Z$ has a term linear in $N$, while the partition function $Z$ contains only even powers of $N$.
As we will discuss in \ref{subsection:largeN}, this is due to the change of topology required by the replica trick.

In figure \ref{figure:shan_fixedN} we plot the Shannon entropy as a function of $A$ at fixed $N$.
While we do not have an analytic prediction for comparison, we will show in the following section that it agrees qualitatively with an analytic approximation of the Shannon entropy.
\subsection{Approximate entropy from the fermion model}
\label{subsection:approximate}

The results of the previous section were obtained from numerical simulations, but we can give an analytic estimate of the $O(N)$ term in the Shannon entropy.
One generically expects the entropy of a system of $N$ fermions to scale linearly with $N$, and this is indeed what we find. 
However, we can go further and obtain a reasonable estimate of the dependence of the coefficient on $A$.
This estimate also provides an upper bound for the Shannon entropy, which gives a further check on the numerical results.

Let us work in the occupation number basis $\vec n =  \{n_h\}_{h \in \mathbb{Z}}$, where $n_h \in \{0,1 \}$.
For each value of $\lambda$, $A$, and $\chi$ there is probability distribution $p(\vec{n})$ for which $S_\text{Shannon}$ is the entropy,
\begin{equation}
S_\text{Shannon} = -\sum_{\vec n} p(\vec n) \log p(\vec n).
\end{equation}
There are nontrivial correlations between the occupation numbers at different sites, but we can obtain a crude approximation by neglecting these correlations.
From the point of view of the lattice fermion model, this is essentially a mean-field approximation where one ignores density-density correlations.
Let $\rho_h$ be the probability that site $h$ is occupied, then the ``one-point entropy''\footnote{This quantity is essentially the one-point entropy defined in Ref.~\cite{Kelly:2013aja}. 
In that context, it was the expectation values of single-trace operators in holographic CFT was held fixed.} is defined as
\begin{equation} \label{S1}
S_{\text{one-point}} = \sum_h [- \rho_h \log \rho_h - (1-\rho_h) \log(1 - \rho_h)].
\end{equation}

In the large $N$ limit, $\rho_h = \rho(h/N)$, where $\rho$ is given by \eqref{rho_weak} or \eqref{rho_strong} depending on the phase.
In this limit the sum becomes an integral, and
\begin{equation}
    S_{\text{one-point}} \sim N \int dh \; \left[ -\rho(h) \log (\rho(h)) - (1- \rho(h)) \log (1-\rho(h)) \right].
\end{equation}
This quantity is plotted in figure \ref{figure:shan_fixedN} along with the Shannon entropy obtained from our simulation, and we find a qualitative agreement. 
The one-point entropy $S_\text{one-point}$ is an upper bound for $S_\text{Shannon}$ as shown in figure \ref{figure:shan_fixedN}.

\subsection{Large $N$ counting and the replica trick}
\label{subsection:largeN}

In section \ref{subsection:numericalresults} we found that the Shannon entropy scales linearly with $N$ to leading order.
From the perspective of the fermion model, this is unsurprising: the entropy of a system of $N$ fermions grows linearly with $N$.
From the string description, where the partition function contains only even powers of $N$, it requires some explanation.
The key point is that when performing the replica trick we analytically continue in the replica index, which requires analytically continuing the Euler characteristic $\chi$ of spacetime.
The powers of $N$ appearing in the partition function depend on $\chi$, so differentiating with respect to $\chi$ disrupts the large $N$ expansion.

To see more explicitly how this works, suppose we were to try to calculate the entanglement entropy for two intervals on a sphere, order by order in $N$, using the Gross-Taylor expansion of the partition function \cite{Gross:1993hu,Gross:1993yt}.
The $n^\text{th}$ replica is a surface with area $A_n = n A$ and Euler characteristic $\chi_n = 4 - 2n$; for $n=1$ we have a sphere, for $n=2$ a torus, etc.
The partition function of replica number $n$ has a large $N$ expansion beginning at order $N^{4 - 2n}$.
This means if we hold the power of $N$ fixed while carrying out the replica trick we will analytically continue a sequence which is identically zero beyond some finite point, leading to nonsensical results.
The resolution is that we should first calculate the entanglement entropy at finite $N$ and take the large $N$ limit only at the end.

\begin{figure}[ht!]
    \centering
    \includegraphics[width=4in]{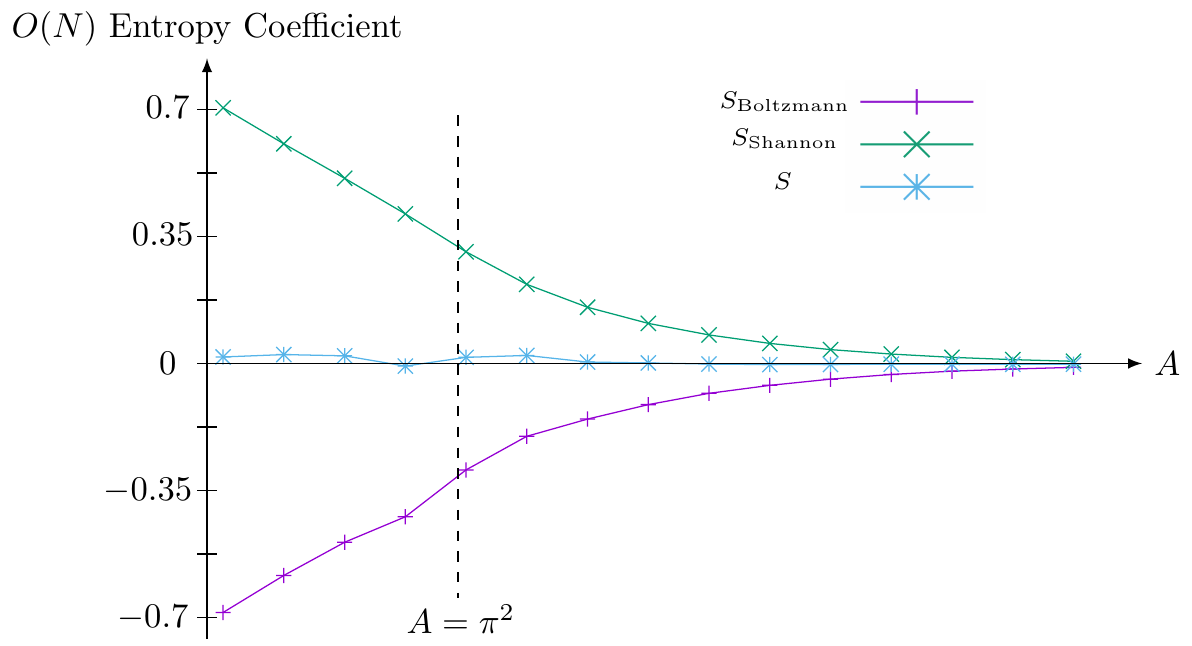}
    \caption{Here we see the cancellation between the $O(N)$ parts of the Shannon and Boltzmann entropies demanded by the closed string expansion.
    The upper (green) curve is the coefficient of $N$ in $S_\text{Shannon}$; the lower (purple) curve is the coefficient of $N$ in $S_\text{Boltzmann}$.
    Their sum is the middle (blue) curve, which shows a behaviour consistent with zero.
    The slight deviation from zero comes from systematic errors in estimating the linear term of the large $N$ expansion of $S_\text{Boltzmann}$ from data gathered at finite $N$. Once again we set $\lambda=1$.}
    \label{figure:cancellation}
\end{figure}

In some special cases, the entanglement entropy does not require analytically continuing to different spacetime topologies.
When $\chi = m$, the total entropy \eqref{totalentropy} can be obtained by differentiating the partition function with respect to $A$.
\begin{equation}
S = (1 - A \partial_A) \log Z(A, \chi).
\end{equation}
This happens when the entangling surface consists of two points on a sphere for which $\chi = m = 2$ or when the entangling surface consists of zero points (i.e. when considering the thermal entropy) on a torus for which $\chi = m = 0$.\footnote{While we have not considered this latter case, the Boltzmann entropy vanishes and we expect the Shannon entropy to take the form of a power series in $1/N^2$ starting from $N^0$.}
These special cases are precisely those where the modular flow is geometric, i.e., where the density matrix $\rho$ associated to a region can be written as $e^{-H}$ where $H$ is a generator of a spacetime symmetry.
In this case, we can express the partition function in the $1/N$ expansion and differentiate term-by-term and so the total entropy must have no term linear in $N$.
Since the Shannon entropy scales linearly with $N$, the only way this can happen is if the positive $O(N)$ term in the Shannon entropy cancels against a subleading negative $O(N)$ term in the Boltzmann entropy.
We test this numerically and find that this is indeed the case: this cancellation is illustrated in figure \ref{figure:cancellation}. 
We note that this cancellation is not at all obvious from the point of view of the fermion model, illustrating the advantage of having multiple dual descriptions of the same physics.

This cancellation is rather special and occurs only for a single-interval entangling surface on the sphere.
If we increase the number of intervals, the Boltzmann entropy scales linearly with the number of points, while the Shannon entropy stays the same.
Thus for two or more intervals we get a negative $O(N)$ term in the entropy.
This is relevant for the mutual information, for example.
If we calculate the mutual information between intervals $\mathbf{A}$ and $\mathbf{B}$,
\begin{equation}
    I(\mathbf{A} : \mathbf{B}) = S(\mathbf{A}) + S(\mathbf{B}) - S(\mathbf{A} \cup \mathbf{B})
\end{equation}
at order $N$ the mutual information comes from the negative term in $S(\mathbf{A} \cup \mathbf{B})$, leading to a positive mutual information of order $N$.

Given the apparent potential hazards of applying the replica trick to the string expansion, one might wonder why the leading $O(N^2)$ term seems to be unaffected.
The answer is that, if we define the analytic continuation of the partition function by the sum \eqref{Z}, the saddle point exists for any $\chi > 0$.
At the saddle point, we can replace $\chi$ derivatives with $A$ derivatives following the argument of section \ref{section:free-energy} and therefore we can calculate term-by-term in the $1/N$ expansion of the partition function.
This argument applies only to the leading $N^2$ term in the saddle point approximation.

\section{Discussion}\label{section:discussion}

We have calculated the entanglement entropy for two-dimensional Yang-Mills theory in the large $N$ limit, focusing on the division of the entropy into what we have called the Boltzmann and Shannon entropies.
The Boltzmann entropy is the expectation value of a local operator on the entangling surface analogous to the Ryu-Takayanagi formula. 
We have shown that this term dominates the large $N$ limit, giving further support for this analogy.

The appearance of a term in the entropy linear in $N$, while evident from the fermion model, is surprising from the point of view of the closed string theory.
We have explained the appearance of such terms as a breakdown of the genus expansion in the string theory when we analytically continue the target space topology.
However we do not have a prescription for computing coefficients of the large $N$ expansion of the entropy from the string expansion.
This would require resumming some class of diagrams across different target space topologies; this is further complicated by the fact that we have to include diagrams with an arbitrary number of interaction vertices.
The resurgence analysis applied to the torus partition function in Ref.~\cite{Okuyama:2018clk} might be useful in this regard.

The analogy with random matrix theory may also provide an explanation for the term linear in $N$\footnote{We thank Sylvain Carozza for pointing out this connection to matrix models.}.
In the weak coupling phase, the fermion positions behave like the eigenvalues of a random matrix, with the factor $\dim (R)^\chi$ playing the role of eigenvalue repulsion in the random matrix model.
The Euler characteristic $\chi$ is then analogous to the parameter $\beta$ of random matrix ensembles, where $\beta = 2$ corresponds to the unitary matrices \cite{Eynard:2015aea}.
At $\beta = 2$, the Feynman diagrams have the topology of oriented surfaces and so the large $N$ expansion contains only even powers of $N$.
But perturbing away from $\beta = 2$ introduces non-oriented surfaces, and, in particular, diagrams with the topology of the real projective plane contribute at linear order in $N$.
This suggests that the string description may involve non-orientable worldsheets; these appear naturally in the string description of two-dimensional Yang-Mills theory with gauge group $\mathrm{O}(N)$ or $\mathrm{Sp}(N)$ \cite{Naculich:1993ve}.

While the fact that the Shannon entropy grows linearly with $N$ is evident from the fermion model, the fact that it cancels (in certain cases) with a subleading term in the Boltzmann entropy is surprising.
A similar cancellation sometimes occurs in quantum field theory at one loop: while the entanglement entropy is always positive, it can cancel against the expectation value of the Wald entropy.
For example, a conformally coupled scalar field theory in four dimensions has an ultraviolet-divergent entanglement entropy proportional to the area of the entangling surface, but this divergence cancels against the Wald entropy term which takes the form $-\xi \langle \phi^2 \rangle$ integrated over the entangling surface \cite{Solodukhin:1995ak}.
It would be interesting to understand whether such cancellations can arise from an underlying closed string description.

Perhaps the most intriguing consequence of this work is the possibility that the breakdown of large $N$ counting implied by the replica trick is a more general phenomenon.
Suppose, for instance, that the same effect is present in $\mathcal{N}=4$ supersymmetric Yang-Mills theory. 
This would imply a breakdown of the closed string genus expansion in the bulk, and corrections to holographic entanglement entropy larger than expected from semiclassical field theory \cite{Faulkner:2013ana}.
This would have potentially drastic implications for recent discussions of the black hole information paradox \cite{Penington:2019npb,Almheiri:2019psf,Almheiri:2019hni} which rely on an interplay between the leading term in the entanglement entropy and its subleading correction.
We believe the possibility of large quantum corrections to black hole thermodynamics is compelling enough to warrant further investigation in other instances of large $N$ gauge/string duality.

\section*{Acknowledgments}
WD acknowledges helpful discussions with Sylvain Carozza, Aron Wall and Gabriel Wong.
NVM acknowledges enlightening discussions with Aurora Ireland, and a helpful discussion about matrix models with Douglas Stanford.
We are all grateful to the Perimeter Institute Undergraduate Summer Program, where this project began. This research was supported in part by Perimeter Institute for Theoretical Physics.
Research at Perimeter Institute is supported by the Government of Canada through the Department of Innovation, Science and Economic Development Canada and by the Province of Ontario through the Ministry of Research, Innovation and Science.

\begin{appendix}
\section{Details on solving the saddle point equation}

Here we provide a more detailed derivation of the solution of the saddle point equations.
For future reference, we use the following notation and conventions for the complete elliptic integrals of the first, second, and third kinds:
\begin{align}
K(m) &= \int_0^1 \frac{dt}{\sqrt{1-t^2}\sqrt{1-m t^2}} , \label{ellipticK} \\
E(m) &= \int_0^1 dt \, \frac{\sqrt{1- m t^2}}{\sqrt{1 - t^2}}, \label{ellipticE} \\
\Pi(n|m) &= \int_0^1 \frac{dt}{(1-nt^2) \sqrt{1-t^2} \sqrt{1 - m t^2}}. \label{ellipticPi}
\end{align}

\subsection{Weak coupling}
We seek to find the resolvent defined in \eqref{R},
\begin{align}
    R(z) = \int_{-a}^a ds \frac{\rho(s)}{z-s}.
\end{align}
Consider the contour integral
\begin{align}
    \oint_{C_\infty}\frac{dw}{2\pi i}\frac{R(w)}{w-z}\frac{g(z)}{g(w)},
\end{align}
with $C_{\infty}$ a contour at infinity. 
We know from \eqref{R} that $R(w) \sim 1/w$ for large $w$, and that $g(w) \sim w$, so the integrand goes like $1/w^3$ and the contour integral vanishes. 
But we also know that this integral is equal to the integrals we carry out on contours around all the singularities of the integrand. Since these are the branch cut and the point $z$, from (\ref{cauchy}) we find
\begin{align}
    R(z) = -\oint_{C_b} \frac{dw}{2\pi i} \frac{R(w)}{w-z}\frac{g(z)}{g(w)},
\end{align}
with $C_b$ chosen to surround the branch cut and exclude $z$. In particular, we choose
\begin{align}
    R(z) = \frac{g(z)}{2\pi i} \left(\int_{-a}^a dh \frac{R(h+i\epsilon)}{h+i\epsilon-z}\frac{1}{g(h+i\epsilon)}-\int_{-a}^a dh \frac{R(h-i\epsilon)}{h-i\epsilon-z}\frac{1}{g(h-i\epsilon)} \right).
\end{align}
The semicircular boundaries of the contour go to 0 as $\epsilon\to0$. Notice that 
\begin{equation}
    g(h\pm i\epsilon) = \pm i \sqrt{a^2-h^2},
\end{equation} 
which leads to
\begin{align}
    R(z) = \frac{-g(z)}{2\pi} \int_{-a}^a dh \frac{R_+(h) + R_-(h)}{(h-z)\sqrt{a^2 - h^2}}.
\end{align}
Here we can use the Sokhotski–Plemelj theorem and our original integral equation \eqref{eq:integralequation} to write this as
\begin{align*}
    R(z) &= -\frac{\lambda A g(z)}{\pi\chi}\int_{-a}^a dh \frac{h}{(h-z)\sqrt{a^2-h^2}} \\
    &= -\frac{\lambda A g(z)}{\pi\chi} \left( \pi - \frac{\pi}{\sqrt{1 - \frac{a^2}{z^2}}} \right) \\
    &= -\frac{\lambda A}{\chi}\left(z\sqrt{1 - \frac{a^2}{z^2}} -z\right)
\end{align*}
This means
\begin{align}
    R(h \pm i \epsilon) = -\frac{\lambda A}{\chi}\left(\pm i \sqrt{a^2 - h^2} - h \right),
\end{align}
which directly leads to $\rho(h)$. 

\subsection{Strong coupling}

Now we want to find the resolvent defined in (\ref{Rtilde}),
\begin{align}
    \tilde{R}(z) = \int_{(-a,-b)\cup (b,a)} dh \frac{\tilde{\rho} (h)}{z - h}.
\end{align}
 With the choice of $g(z)$ defined in \eqref{gz} we have
\begin{align}
    \oint_{C_\infty} \frac{dw}{2 \pi i} \frac{\tilde{R}(w)}{w-z} \frac{g(z)}{g(w)}=0,
\end{align}
where $C_{\infty}$ is a contour at infinity. Like before, this integral must be equal to the integrals around all singularities of the integrand, so
\begin{align}
    \tilde{R}(z) = -\oint_{C_b} \frac{dw}{2 \pi i} \frac{\tilde{R}(w)}{w-z} \frac{g(z)}{g(w)},
\end{align}
where $C_b$ is a contour that encircles both branch cuts of $g(w)$ counterclockwise.
If we choose this contour to be infinitesimally close to each branch, there is no contribution from the endpoints of the branch cuts and we have
\begin{align}
\begin{split}
    \tilde{R}(z) = \frac{g(z)}{2\pi i} &\left( \int_{-a}^{-b} dh \frac{\tilde{R}(h+i\epsilon)}{(h+i\epsilon-z) g(h+i\epsilon)} -
    \int_{-a}^{-b} dh \frac{\tilde{R}(h-i\epsilon)}{(h-i\epsilon-z)g(h-i\epsilon)} \right.
    \\ &\left. +\int_{b}^{a} dh \frac{\tilde{R}(h+i\epsilon)}{(h+i\epsilon-z) g(h+i\epsilon)} -
    \int_{b}^{a} dh \frac{\tilde{R}(h-i\epsilon)}{(h-i\epsilon-z)g(h-i\epsilon)} \right).
    \end{split}
\end{align}
Now we use the fact that in the region $b < |h| < a$, $g(h-i\epsilon) = -g(h+i \epsilon)$.
This flips the relative sign of the upper and lower branch, so now the values of $R$ above and below the cut are added rather than subtracted:
\begin{align}
\tilde{R}(z) = \frac{g(z)}{2\pi i} \int_{(-a,-b) \cup (b,a)} dh \frac{\tilde{R}(h+i\epsilon) + \tilde{R}(h - i \epsilon)}{(h-z) g(h + i \epsilon)}.
\end{align}
Applying the Sokhotski-Plemelj formula and equation \eqref{strong}, we can eliminate $\tilde R$ from the right-hand side.
We can then reverse the previous process to go back to contour integrals around the branch cuts:
\begin{align}
     \tilde{R}(z) &= \frac{g(z)}{2\pi i} \int_{(-a,-b) \cup (b,a)} dh \frac{2\frac{\lambda A}{\chi} h +2\log\frac{h-b}{h+b}}{(h-z) g(h + i \epsilon)} \\
     &= \frac{g(z)}{2\pi i} \left( \int_{(-a,-b) \cup (b,a)} dh \frac{\frac{\lambda A}{\chi} h + \log\frac{h-b}{h+b}}{(h-z) g(h + i \epsilon)} -  \int_{(-a,-b) \cup (b,a)} dh \frac{\frac{\lambda A}{\chi} h + \log\frac{h-b}{h+b}}{(h-z) g(h - i \epsilon)} \right).
\end{align}
As the numerator has no branch cuts in the region of integration, this is equivalent to the contour integral around each branch cut,
\begin{align}
    \tilde{R}(z) = \frac{-g(z)}{2\pi i} \oint_{C_b} dw\frac{\frac{\lambda A}{\chi}w +\log\frac{w-b}{w+b}}{(w-z)g(w)}.
\end{align}
Now we can consider
\begin{align}
    \oint_{C_\infty} dw\frac{\frac{\lambda A}{\chi}w +\log\frac{w-b}{w+b}}{(w-z)g(w)},
\end{align}
which is zero, but also equal to the sum of contour integrals around all singularities of the integrand. 
So we can again deform the contour:
\begin{align}
    \tilde{R}(z) = \frac{g(z)}{2\pi i} \left(\oint_{C_z} + \oint_{C_l}\right)dw \frac{\frac{\lambda A}{\chi}w +\log\frac{w-b}{w+b}}{(w-z)g(w)},
\end{align}
where $C_l$ is a contour encirculing the branch cut of the logarithm counterclockwise. 
The $C_z$ integral is straightforward thanks to the residue theorem, and the $C_l$ integral is also simple because the jump of the logarithm across the branch cut is constant,
\begin{align}
\tilde{R}(z) &= \frac{\lambda A}{\chi}z + \log\frac{z-b}{z+b} -g(z)\int_{-b}^b \frac{dh}{(h-z)g(h)}.
\end{align}
We can evaluate the integral over $h$ by multiplying the numerator and denominator by $(h+z)$, using parity arguments and a change of variables, to arrive at the definition of the complete elliptic integral of the third kind \eqref{ellipticPi},
\begin{align}
\tilde{R}(z) = \frac{\lambda A}{\chi}z + \log\frac{z-b}{z+b}  -\frac{2 g(z)}{az} \; \Pi\left(\frac{b^2}{z^2}\middle| \frac{b^2}{a^2}\right).
\end{align}
The density $\rho(h)$ then follows from the Sokhostki-Plemelj theorem and 
\begin{equation}
    R(h \pm i \epsilon) = \frac{\lambda A}{\chi} h \mp \text{sgn}(h)\frac{2 i }{ah}\sqrt{a^2-h^2}\sqrt{h^2 - b^2} \; \Pi\left(\frac{b^2}{h^2}\middle| \frac{b^2}{a^2}\right).
\end{equation}
Note that the $\log$ term drops out when going from $\tilde R$ to $R$.
\end{appendix}

\bibliographystyle{utphys}
\bibliography{2dym}

\end{document}